\documentclass[12pt]{article}
\usepackage{graphicx}

\usepackage{amsmath}
\usepackage{amssymb}

\newcommand\pubnumber{}
\newcommand\pubdate{\today}

\def\liege{IFPA, Dep. AGO, Universit\'e de Li\`ege, Bat B5, Sart-Tilman B-4000
  Li\`ege 1, Belgium}

\def\Title#1{\begin{center} {\Large #1 } \end{center}}
\def\Author#1{\begin{center}{ \sc #1} \end{center}}
\def\Address#1{\begin{center}{ \it #1} \end{center}}

\newcommand\pubblock{\rightline{\begin{tabular}{l} \pubnumber\\
         \pubdate  \end{tabular}}}
\newenvironment{Abstract}{\begin{quotation}  }{\end{quotation}}
\newenvironment{Presented}{\begin{quotation} \begin{center} 
             PRESENTED AT\end{center}\bigskip 
      \begin{center}\begin{large}}{\end{large}\end{center} \end{quotation}}
\def\Acknowledgements{\bigskip  \bigskip \begin{center} \begin{large}
             \bf ACKNOWLEDGEMENTS \end{large}\end{center}}

\def\beq{\begin{equation}}
\def\eeq#1{\label{#1}\end{equation}}
\def\eeqn{\end{equation}}

\def\beqa{\begin{eqnarray}}
\def\eeqa#1{\label{#1}\end{eqnarray}}
\def\eeqan{\end{eqnarray}}

\let\bar=\overbar

\def\Dslash{\not{\hbox{\kern-4pt $D$}}}
\def\dslash{\not{\hbox{\kern-2pt $\del$}}}

\def\msb{{\bar{\ssstyle M \kern -1pt S}}}

\begin{document}
\begin{titlepage}
\pubblock

\vfill
\Title{Theory and phenomenology of lepton flavor violation}
\vfill
\Author{Avelino Vicente}
\Address{\liege}
\vfill
\begin{Abstract}
The field of lepton flavor violation will live an era of unprecedented developments in the near future, with dedicated experiments in different fronts. The observation of a flavor violating process involving charged leptons would be a clear evidence of physics beyond the Standard Model, thus motivating the great effort in this direction. Furthermore, in case a positive signal is found, a proper theoretical understanding of the lepton flavor anatomy of a given model would become necessary. Here I briefly review the current situation, emphasizing the most relevant theoretical and phenomenological aspects of several processes. Finally, I discuss two topics that have received some attention recently: lepton flavor violation in low-scale seesaw models and lepton flavor violating Higgs decays.
\end{Abstract}
\vfill
\begin{Presented}
8th International Workshop on the CKM Unitarity Triangle (CKM 2014), Vienna, Austria, September 8-12, 2014
\end{Presented}
\vfill
\end{titlepage}
\def\thefootnote{\fnsymbol{footnote}}
\setcounter{footnote}{0}

\section{Introduction}
\label{sec:intro}

The field of lepton flavor violation (LFV) is about to begin a golden
era, with great expectations in several experimental
projects~\footnote{See Section 1 of Ref. \cite{Abada:2014kba} and
  references therein for a complete review of the current experimental
  situation and the future prospects.}. In the coming years, many
collaborations will join the search for LFV, currently led by the
popular MEG experiment. These new experiments will look for LFV in
channels that not only include radiative lepton decays, but also
3-body lepton decays (such as $\mu \to 3 \, e$), $\mu-e$ conversion in
nuclei or LFV in high-energy colliders. With these great perspectives,
we may be able to extend our knowledge on the physics beyond the
Standard Model (SM) or, at least, to significantly improve the current
bounds.

On general grounds, one expects large LFV effects if new physics
exists close to the electroweak scale. In fact, most popular models
predict large LFV rates. This is easily understood by simple
considerations based on effective field theory. Let us consider the
dim-6 operator
\begin{equation}
\mathcal{O}_{e \mu} = \frac{c_{e \mu}}{\Lambda^2} \, \bar \mu e \bar e e \, ,
\end{equation}
induced by some heavy degrees of freedom with masses of the order of
$\Lambda$. It violates the electron and muon flavors, thus inducing
processes such as $\mu \to 3 \, e$. Using the current bounds on this
process one finds that the condition $\Lambda / \sqrt{c_{e \mu}} >
100$ TeV must be satisfied. Therefore, if new physics capable of
inducing the operator $\mathcal{O}_{e \mu}$ is found at the TeV scale,
some suppression mechanism must be introduced in order to satisfy the
current LFV bounds. This, together with the promising experimental
perspectives, makes LFV an interesting road in the search for new
physics, complementary to the direct path based on high-energy
colliders.

In case a positive observation in one or several experiments is made,
the correct interpretation of the results in a given model will
definitely require a detailed understanding of its LFV anatomy. This
theoretical effort should be ambitious. In addition to detailed
computations, patterns and correlations must be properly identified in
order to be able to extract as much information as possible. Only by
combining these \emph{tests}, typically valid for general classes of
models, one can investigate the physics responsible for LFV.

An example of one of these patterns is the so-called \emph{dipole
  dominance}. In many popular models, the operators with the dominant
contributions to LFV processes are dipole operators induced by photon
exchange. This is for example the case of the Minimal Supersymmetric
Standard Model (MSSM). In this type of scenarios there is a very
strong correlation between radiative lepton decays and the
corresponding 3-body lepton decays,
\begin{equation}
\frac{\text{BR}(\ell_i \to 3 \, \ell_j)}{\text{BR}(\ell_i\to \ell_j \gamma)} =  \frac{\alpha}{3\pi}\left(\log\frac{m_{\ell_i}^2}{m_{\ell_j}^2}-\frac{11}{4}\right) \, ,
\end{equation}
which allow to obtain a clear hierarchy between LFV observables,
$\text{BR}(\ell_i\to \ell_j \gamma) \gg \text{BR}(\ell_i \to 3 \,
\ell_j)$. An analogous relation can be found for other LFV processes,
like $\mu-e$ conversion in nuclei, that are also suppressed with
respect to the radiative muon decay. The violation of these
hierarchies would be a clear signal of a departure from these standard
scenarios, and thus they should be considered as powerful experimental
tests.

In the following we are going to discuss two specific topics: LFV in
low-scale seesaw models and Higgs LFV decays.

\section{LFV in low-scale seesaw models}
\label{sec:lowscale}

Low-scale seesaw models offer an interesting alternative to the usual
high-energy realizations of the seesaw mechanism. Instead of a large
mass scale, low-scale seesaw models rely on the violation of lepton
number by a small parameter. This allows one to explain the smallness
of neutrino masses and, simulaneously, have right-handed (RH)
neutrinos at the TeV scale (or below), thus contributing to LFV
processes.

In the Inverse Seesaw \cite{Mohapatra:1986bd}, the SM particle content
is extended by $n_R$ generations of RH neutrinos $\nu_R$ and
$n_X$ generations of singlet fermions $X$, both with lepton number
$L=+1$. In the following we will assume $n_R = n_X = 3$, although more
minimal models are also possible \cite{Abada:2014vea}. The Lagrangian
has the form
\begin{equation}
 \mathcal{L}_\mathrm{ISS} = \mathcal{L}_\mathrm{SM} - Y^{ij}_\nu \overline{\nu_{Ri}} \widetilde{H}^\dagger L_{j}-M_R^{ij}\overline{\nu_{Ri}} X_j-\frac{1}{2}\mu_{X}^{ij}\overline{X_{i}^C} X_{j}+ h.c. \, ,
\label{ISS}
\end{equation}
where a sum over $i, j = 1, 2, 3$ is assumed. Here
$\mathcal{L}_\mathrm{SM}$ is the SM Lagrangian, $Y_\nu$ are the
neutrino Yukawa couplings, $M_R$ is a complex Dirac mass matrix for
the fermion singlets and $\mu_X$ is a complex symmetric Majorana mass
matrix that violates lepton number by two units. The supersymmetric
(SUSY) extension of this model is simply obtained by promoting all
fields to superfields. Furthermore, we note that $\mu_X$ is naturally
small, in the sense of 't Hooft~\cite{'tHooft:1979bh}, since in the
limit $\mu_X \to 0$ lepton number is restored.

After electroweak symmetry breaking, in the basis
$(\nu_L\,,\;\nu_R^C\,,\;X)$, the $9\times 9$ neutrino mass matrix is
given by
\begin{equation}
 M_{\mathrm{ISS}}=\left(\begin{array}{c c c}
0 & m_D^T & 0 \\
m_D & 0 & M_R \\ 
0 & M_R^T & \mu_X \end{array}\right)\,.\label{eq:ISSmatrix}
\end{equation}
where $m_D = \frac{1}{\sqrt{2}} Y_\nu v$ and $v/\sqrt{2}$ is the
vacuum expectation value (VEV) of the Higgs field. Under the
assumption $\mu_X \ll m_D \ll M_R$, the mass matrix $M_{\mathrm{ISS}}$
can be block-diagonalized to give the effective mass matrix for the
light neutrinos $m_{\mathrm{light}}\simeq m_D^T {M_R^T}^{-1} \mu_X
M_R^{-1} m_D$, whereas the heavy quasi-Dirac neutrinos have masses
corresponding approximately to the entries of $M_R$. Since the light
neutrino masses are proportional to $\mu_X$, the smallness of this
parameter can be used to accommodate the values measured in neutrino
oscillation experiments while having $M_R \sim \text{TeV}$. This has
important consequences for LFV.

Early works on LFV in models with light RH neutrinos
\cite{Bernabeu:1987gr,Ilakovac:1994kj,Deppisch:2004fa,Deppisch:2005zm}
already pointed out the existence of large enhancements in the LFV
rates with respect to those found in high-scale models. More recently,
dominant RH neutrino contributions have been found in box diagrams
\cite{Ilakovac:2009jf,Alonso:2012ji,Dinh:2012bp,Ilakovac:2012sh}, as
well as in the usual photon penguin diagrams
\cite{Dev:2013oxa}. Finally, the first complete LFV study including
all contributions in the non-supersymmetric as well as in the
supersymmetric version of the inverse seesaw was presented in
\cite{Abada:2014kba}. The calculation of the LFV observables was done
with {\tt FlavorKit} \cite{Porod:2014xia}, a computer tool that allows
for an automatized analytical and numerical computation of flavor
observables. Thanks to this tool, it was possible to evaluate several
hundreds of Feynman diagrams and obtain complete expressions for the
LFV observables of interest.

\begin{figure}[tb]
\centering
\includegraphics[width=0.49\linewidth]{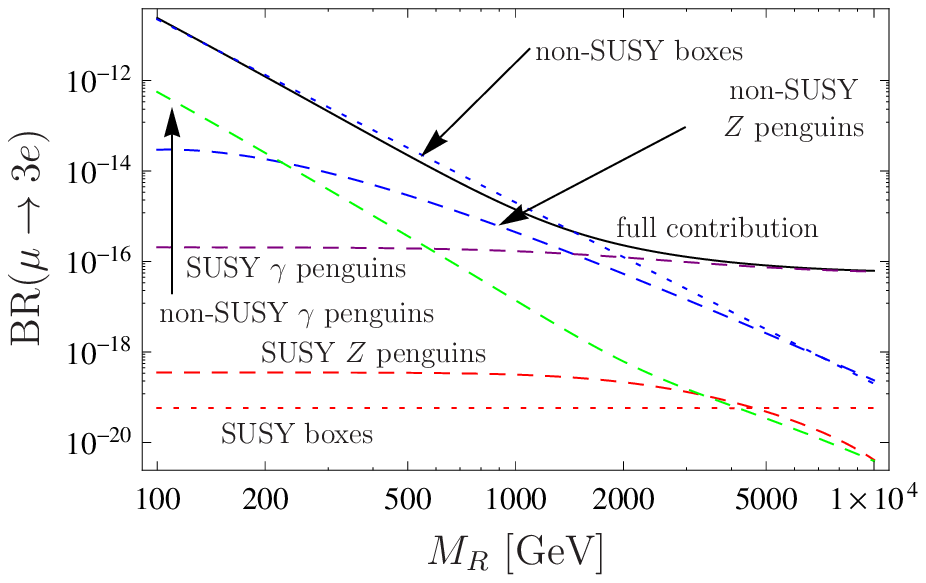}
\includegraphics[width=0.49\linewidth]{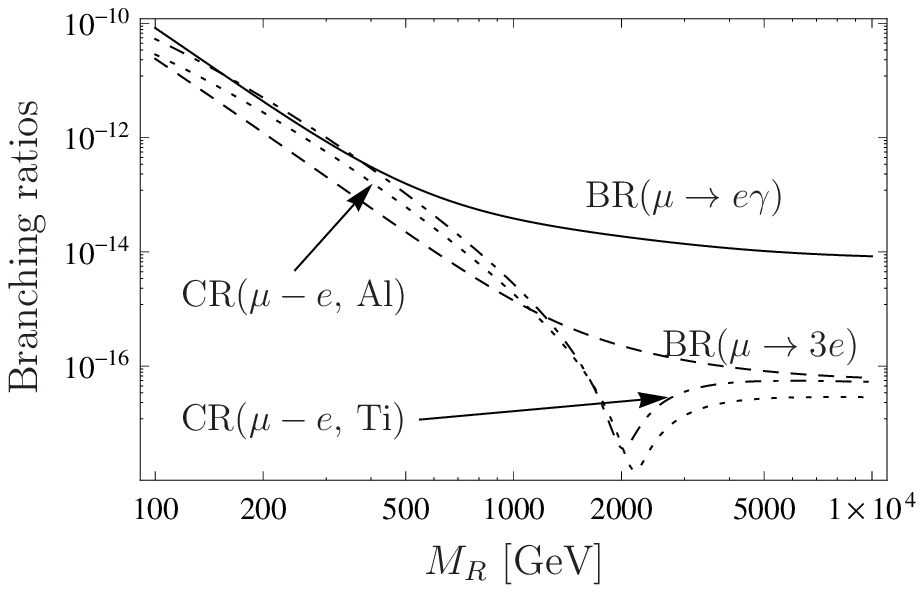}
\caption{On the left-hand side, BR($\mu\to 3 e$) as a function of
  $M_R$. The continuous black line represents the total branching
  ratio, whereas individual contributions are shown with dashed
  lines. On the right-hand side, BR($\mu\to e \gamma$), BR($\mu\to 3
  e$), $\mu-e$ conversion rates in $\mathrm{Ti}$ and $\mathrm{Al}$ as
  a function of $M_R$. Figure from \cite{Abada:2014kba}.}
\label{fig:lowscale}
\end{figure}

As discussed in Sec. \ref{sec:intro}, many popular models have a
dipole dominated LFV phenomenology. In this case, one expects specific
hierarchies among LFV observables. In contrast, in the presence of
light RH neutrinos the dipole dominance is broken. This is shown in
Fig. \ref{fig:lowscale}, obtained from Ref. \cite{Abada:2014kba}. On
the left, individual contributions to BR($\mu\to 3 e$) are shown as a
function of $M_R$. Regarding the supersymmetric parameters, they were
obtained at low energies using renormalization group running from the
grand unification scale, where universal (and flavor-blind) CMSSM-like
boundary conditions were imposed. All universal SUSY breaking
parameters ($m_0$, $M_{1/2}$ and $A_0$) were fixed to $1$ TeV. One
finds that for low $M_R$ the non-SUSY contributions induced by the RH
neutrinos dominate the total branching ratio. In particular, the
non-SUSY boxes, as well as the non-SUSY $\gamma$- and $Z$-penguins,
give the largest contributions. This confirms some partial results in
previous works. The same choice of parameters was made on the
right-side of Fig. \ref{fig:lowscale}, where several LFV observables
are shown. As expected, for low $M_R$ they can have similar rates,
thus breaking the usual hierarchies found for the dipole dominance
scenario.

\section{Higgs LFV decays}
\label{sec:HiggsLFV}

The long-awaited discovery of the Higgs boson should not be seen as
the end of the way, but as the beginning of an era in particle
physics. A new particle always implies new measurements, and the Higgs
boson properties and decay modes might hide very valuable
information. Currently, the open question is whether the discovered
state corresponds to the \emph{standard} Higgs boson. If this is not
the case, and a deviation from the SM expectation is found, we may be
able to get a hint about new physics not far from the electroweak
scale.

Recently, the possibility of LFV Higgs decays
\cite{Pilaftsis:1992st,DiazCruz:1999xe} has attracted some
attention. In addition to model independent studies on LFV Higgs
couplings and their phenomenological impact
\cite{Goudelis:2011un,Blankenburg:2012ex,Harnik:2012pb,Celis:2013xja},
several groups have searched for specific models capable of producing
sizeable LFV Higgs decays with observable rates at the LHC. With
$\sqrt{s} = 8$ TeV and $20$ fb$^{-1}$, the LHC is estimated to be
sensitive to $\text{BR}(h \to \ell_i \ell_j) \gtrsim 10^{-3}$
\cite{Davidson:2012ds}, which implies a very strong requirement on a
model that aims at observable LFV Higgs decays. In fact, most models
fail, as those large LFV Higgs rates would be in conflict with bounds
from radiative lepton decays ($\ell_i \to \ell_j \gamma$).

\begin{figure}[tb]
\centering
\includegraphics[width=0.85\linewidth]{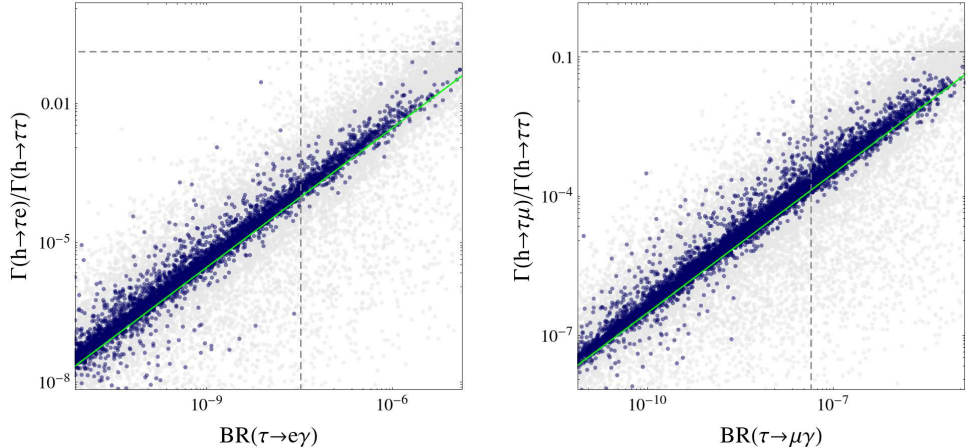}
\caption{Correlation between LFV Higgs decays and radiative lepton
  decays. Gray points are ruled out by LHC direct searches. The
  experimentally allowed region is left of and below the dashed
  lines. Figure from \cite{Falkowski:2013jya}.}
\label{fig:vectorlike}
\end{figure}

Let us consider an example. In \cite{Falkowski:2013jya}, a generic
model with additional vector-like leptons was investigated. The
relevant Lagrangian terms are given by
\begin{eqnarray}
{\cal L}_{F,c} &=& - M \left( \bar L C_L  L + \overline{\tilde  E} C_R \tilde E   \right) 
- \left( \bar L_L Y \tilde E_R H +  \bar L_R \tilde Y \tilde E_L  H + {\rm h.c.} \right) \, , \\
{\cal L}_\text{mix} &=&  M \left ( \bar l_L   \lambda_l L_R  +   \overline{\tilde  E_L} \lambda_e   e_R  \right ) +  {\rm h.c.} \, .
\end{eqnarray}
Here we have introduced the usual $3$ generations of chiral leptons,
$l^i_L=(\nu_L^i,e_L^i)$, $e_R^i$, $i=1\dots 3$, and $3$ generations of
vector-like leptons $L^i = (N^i,E^i)$, $\tilde E^i$, transforming as
$2_{-1/2}$ and $1_{-1}$ under the electroweak gauge group.  $C_L$,
$C_R$, $Y$, $\tilde Y$ and $\lambda_i$ are $3 \times 3$ matrices in
generation space and a common scale $M$, the vector-like mass, has
been isolated. The mixing between the SM leptons and the exotic
states, together with the coupling between the exotic leptons and the
Higgs boson, leads to effective off-diagonal Higgs couplings to a pair
of charged leptons,
\begin{equation}
{\cal L}_\text{eff} =  
- \frac{h}{ \sqrt 2} \bar e_L   c_{\rm eff } e_R    +  {\rm h.c.}  
\qquad
c_{\rm eff} =  Y_{\rm eff} + \frac{v^2}{M^2} \lambda_l C_L^{-1} Y C_R^{-1} \tilde Y C_L^{-1} Y C_R^{-1}\lambda_e  \, ,
\label{eq:ceff}
\end{equation}
where $Y_{\rm eff}$ is flavor diagonal.  However, the same effective
couplings, $c_{\rm eff}$, contribute to radiative lepton decays,
giving rise to a strong correlation between $\text{BR}(h \to \ell_i
\ell_j)$ and $\text{BR}(\ell_i \to \ell_j \gamma)$. This is shown in
Fig. \ref{fig:vectorlike}. When the current bounds on $\text{BR}(\tau
\to e \gamma)$ and $\text{BR}(\tau \to \mu \gamma)$ are used, this
correlation translates into Higgs LFV branching ratios below
$10^{-5}$, below the LHC reach.

Similar perspectives are found in other scenarios
\cite{Arhrib:2012mg,Arana-Catania:2013xma,Arganda:2014dta}. In
contrast, as a positive note, models with extended scalar sectors can
in principle lead to observable LHV Higgs decays. The simplest of
these frameworks is the Type-III Two Higgs Doublet Model
(2HDM)~\footnote{More complicated multi-Higgs models have been
  examined too
  \cite{Bhattacharyya:2010hp,Bhattacharyya:2012ze,Arroyo:2013kaa,Campos:2014zaa}.}.
Already suggested in previous works
\cite{Davidson:2010xv,Harnik:2012pb,Kopp:2014rva}, this has been
recently shown explicitly in \cite{Sierra:2014nqa}, where the most
relevant constraints on the Type-III 2HDM parameter space were taken
into account.

\section{Summary}
\label{sec:summary}

This mini-review on lepton flavor violation discusses some general
aspects of LFV, emphasizing those that can be used to get a hint on
the underlying physics. In particular, the search for correlations and
hierarchies among LFV observables have been shown to be useful tests
in many scenarios. These patterns are not only possible, but expected
features of general classes of models.

We have briefly reviewed two specific topics that illustrate this
program. First, we have considered LFV in low-scale seesaw models,
where the dominance of dipole operators gets broken due to the
presence of light right-handed neutrinos. This leads to a richer
phenomenology, with several observables with similar rates. Then we
have discussed Higgs LFV decays, reviewing some proposals in the
literature that fail to give observable rates at the LHC due to other
flavor bounds. The most general 2HDM (the so-called Type-III version)
can however lead to large $h \to \ell_i \ell_j$ branching ratios, thus
becoming the perfect scenario for Higgs induced flavor
effects~\footnote{We have refrained from discussing the recent $2.5
  \sigma$ excess found by the CMS collaboration in the $h \to \tau
  \mu$ channel \cite{CMS:2014hha}. This intriguing excess would
  translate into a largish BR$(h\to\tau\mu) = \left(
  0.89_{-0.37}^{+0.40} \right)$\%, thus favoring an explanation based
  on the Type-III 2HDM \cite{Sierra:2014nqa}.}.

\Acknowledgements
I thank the organizers of the CKM 2014 conference in Vienna for the
exciting scientific atmosphere in the meeting as well as the rest of
participants for the rich discussions during the conference.

\end{document}